\documentstyle[12pt]{article}
\newcommand{\eqnum}[1]{\setcounter{equation}{#1}}

\newcommand{\non}{\nonumber}
\newcommand{\beq}{\begin{equation}}
\newcommand{\eeq}{\end{equation}}
\newcommand{\barr}{\begin{eqnarray}}
\newcommand{\earr}{\end{eqnarray}}

\newcommand{\square}{\kern1pt\vbox{\hrule height 1.2pt\hbox
{\vrule width1.2pt\hskip 3pt \vbox{\vskip
6pt}\hskip 3pt\vrule width 0.6pt}\hrule height 0.6pt}\kern1pt}

\begin{document}

\renewcommand{\Large}{\large}
\renewcommand{\huge}{\large}
\bibliographystyle{unsrt}

\baselineskip = 14pt

\begin{titlepage}
\baselineskip .15in
%\begin{flushright}
%WU-AP/45/95
%\end{flushright}

~\\

\vskip 1.5cm
\begin{center}
{\bf

\vskip 1.5cm
{\large\bf Chaos in  Static Axisymmetric Spacetimes I :
Vacuum Case}

}\vskip .8in

Yasuhide {\sc Sota}$^{(a)}$, Shingo {\sc  Suzuki}
$^{(b)}$and  Kei-ichi {\sc Maeda}$^{(c)}$\\[.5em] 
{\em Department of Physics, Waseda
University, Shinjuku-ku, Tokyo 169, Japan}
\end{center}
\vfill
%%%%%%%%%%%%%%%%%%%%%%%%%%%%%%%%%%%%%%%%%%%%%%%%%%%%%%%
 \begin{abstract}
We study  the motion of test particle in static
axisymmetric vacuum  spacetimes and discuss two criteria
for strong chaos to occur:
 (1) a local instability measured by the Weyl curvature, 
and (2)  a tangle of a homoclinic orbit, which is
closely  related to an unstable periodic orbit in general
relativity.  We analyze several static axisymmetric
spacetimes and find that the
 first criterion is a sufficient condition for chaos, at least qualitatively.
Although some test particles which do  not
satisfy the first criterion show   
chaotic behavior in some spacetimes, these can be
accounted for the second criterion.
\end{abstract}
%%%%%%%%%%%%%%%%%%%%%%%%%%%%%%%%%%%%%%%%%%%%%%%%%%%%%%%%%%

\vfill
\begin{center}
October, 1995
\end{center}
\vfill
(a)~~electronic mail :  64L511@cfi.waseda.ac.jp\\
(b)~~electronic mail :  64L070@cfi.waseda.ac.jp\\
(c)~~electronic mail :  maeda@cfi.waseda.ac.jp\\
\end{titlepage}

%
%
%
%
%   11    11    11    11    11    11    [Intro]
%
%
%23456789 123456789 123456789 123456789 123456789 123456789
\section{Introduction} 
Chaos has become one of the most important ideas used to
understand various non-linear  phenomena in nature.   We
know many features of  chaos in the Newtonian dynamics.  
However, we do not know, so far, so much  about those in
general relativity (GR).   If gravity is strong, we have 
to use the Einstein's theory of gravitation. We may find
new types of  chaotic behavior in strong gravitational
fields, which do not appear  in Newtonian dynamics.  
Although a chaotic behavior of the non-linear gravitational
field itself may be much more interesting to study, there
may be a fundamental difficulty to discuss its chaotic
dynamics 
 even in the well-studied Bianchi IX model, due to 
a gauge invariance, in particular a choice of  time
coordinate, in GR\cite{Rugh}. Before we  devote ourselves
to studying such a subject, we need to 
 analyze  problems with a clear setting to better
understand chaos in GR.  One such approach is to study test
particle motion in a fixed 
 curved background
\cite{Conto}--\cite{Yurt}.
We have an appropriate time coordinate, the proper time of
a test particle, which is a natural invariant
time\cite{Sota}  or an  infinite observer's time, which has
an invariant meaning because of the existence of a
time-like Killing vector\cite{Cornish}.

In the present  paper,   we study 
test  particle motion in static
axisymmetric vacuum spacetimes and search for  a generic
criterion for chaos to occur.  In Newtonian dynamics, a
connection between chaos and a local instability of a test
particle in a given potential $V$ has been intensively
discussed in two ways. One way is to study local
instability   measured by  eigenvalues of the tidal-force
matrix, which is derived from  linearization of the
equations of motion, i.e., 
\beq
 \frac{d^2 n^i}{dt^2}=- V^i_{~j} n^j , 
\label{tidal}
\eeq
where
$n^i$ is a separation vector between two close orbits
 and $ V^i_{~j}
\equiv \partial^i\partial_j V$  is the tidal-force 
matrix\cite{Toda}{\renewcommand{\thefootnote}
{\fnsymbol{footnote}}\footnote{The Greek and Latin indices 
are used as those of 4-dimensional Lorentzian coordinates
 and  of the 3-dimensional spatial coordinates, respectively.}. 
When the eigenvalues of 
 $V^i_{~j}$ are  all negative at some point, the orbit
becomes locally  unstable in any direction. Then there is a
possibility that such a local instability determines  a
global chaotic behavior of a  test particle.
The other way to analyze a local instability is to use the
curvature of a fictitious space\cite{Arnold}.
 In Newtonian
dynamics,  the orbit of a test particle in a given
potential $V$ does  not follow a geodesic except for a free
particle, which is not interesting in this context. The
equations of motion is 
\beq
\frac{d^2 x^i}{dt^2}=- \partial^i V.  
\label{motion}
\eeq
If the potential $V$ does not directly depend on the time
parameter $t$, the energy of particle $E$ is conserved and the particle  moves inside the classically-allowed region $D=\{ x|V(x)\leq E \}$.
Then we make a conformal transformation of a three
dimensional metric,
\beq \tilde{f}_{ij}=2[E-V(x)]f_{ij}, 
\label{metric}
\eeq
where $f_{ij}$ is the metric of physical flat space.
Using this conformal transformation and transforming the
time parameter from $t$ to $\tilde{t}$ by
\beq
d\tilde{t}=2 [E-V(x)]dt, 
\label{paramet}
\eeq
we obtain an affine parameterized geodesic equation in a
fictitious space with metric $\tilde{f}_{ij}$.
The equation for a deviation vector  $\tilde{\bf{n}}$ orthogonal
to the velocity $\tilde{\bf{v}}$
in the fictitious space is
\beq
 \frac{d^2 \tilde{\bf{n}}}{d\tilde{t}^2}
=-\nabla_{\tilde{\bf{n}}} \tilde{U}, 
\label{tidal2}
\eeq
where  a potential $\tilde{U}$ is given by $\tilde{\bf{v}}$
and $\tilde{\bf{n}}$, such as
\beq
\tilde{U}=\frac{1}{2}\tilde{K}| \tilde{\bf{v}}|^2 | \tilde{\bf{n}}|^2.
\label{tidal3}
\eeq
$\tilde{K}$ is 
a sectional curvature spanned
by $\tilde{\bf{v}}$  and $\tilde{\bf{n}}$ in the fictitious space
and is defined as  \beq
\tilde{K}(\tilde{\bf{v}},\tilde{\bf{n}}) =
\tilde{R}_{ijkl}\tilde{n}^i \tilde{v}^j \tilde{n}^k\tilde{v}^l/
| \tilde{\bf{v}}|^2 | \tilde{\bf{n}}|^2,
\label{sec_curv}
\eeq
where $\tilde{R}_{ijkl}$ is the Riemann tensor given by the 
metric $\tilde{f}_{ij}$ in the fictitious space.
  Since $\tilde{\bf{n}}$=0 is an unstable point of the 
potential $\tilde{U}$ if $\tilde{K}<0$,
a local instability is determined by  a sign of the 
sectional curvature
$\tilde{K}(\tilde{\bf{v}},\tilde{\bf{n}})$.

It has been shown that the
first way to measure a local instability by $V^i_j$ does not always
predict the chaotic motion of a test particle. Some
counter-examples are
known\cite{Eck}--\cite{Biesiada},
where it was pointed out that this non-correlation is
caused by an  adiabatic property in 
Eq.(\ref{tidal})\cite{Cerjan,Biesiada}. In fact, the
tidal-matrix does not have any information about a
particle's velocity, that is, how the orbit passes through
each point in a phase space. Hence,  it seems to be
insufficient to determine the chaotic behavior of each
orbit only from the tidal-force matrix.  However, see the
discussion in \S 3.3. for a relativistic case.

 On the other hand, for the second criterion of
local instability, as was shown in the above,   the local
instability is determined by a sectional curvature of a
2-dimensional plane spanned by the velocity and deviation
vector of the orbit. In this case, the local instability is
determined not only by the position of a particle in
configuration space  but also by its  velocity. It seems to
be a better criterion to determine chaotic behavior. In
fact,  it is known that in the case that   sectional
curvatures of the particle orbit  in a compact
region are all negative, the geodesic becomes
chaotic\cite{Arnold}. If a configuration space is two
dimensional, because  the  sectional curvature is always
equal to the scalar curvature, a negative scalar curvature
always leads to  chaos.

Such a correlation
between the scalar curvature and chaos has been
well-examined  in Newtonian dynamics. For example, in the
two dimensional Kepler problem, all  bounded orbits pass
through only the positive curvature region\cite{Ong}, which
is consistent with the integrability of the system.

  However, it turns out that the second method
 as well as the first one does not
always predict an occurrence of chaos even in two
dimensional case. For example,  in the Henon-Heiles system,
which is a model of a star moving in a galaxy, we  cannot
predict the change of the particle behavior from regular to
chaotic by the sign of scalar curvature because  it is 
always positive although chaos occurs in some
situations\cite{Aizawa}. Furthermore there are some
examples in which  stable periodic orbits exist even in
the  region of negative scalar
curvature\cite{Biesiada,Aizawa}. This failure seems to stem
from the following two main obstacles:\\ (1)  The scalar
curvature diverges on   the boundary of the allowed region
of a particle in Riemannian case{\renewcommand{\thefootnote}
{\fnsymbol{footnote}}\footnote[1]{Here Riemannian and pseudo-Riemannian denote the manifold whose signature is $(+,+,+,\dots)$ and $(-,+,+,\dots)$, respectively.}.

 This singularity is inevitable as long as we
use a conformal transformation (\ref{metric}). In pseudo-Riemannian
case, although a orbit before transformation can cross the boundary of $D$, the curvature after the transformation again diverges on the boundary.
So we cannot follow the orbit near the boundary in the fictitious space for Riemannian and pseudo-Riemannian cases.
Such a problem
has been pointed out by several
authors\cite{Biesiada,Burd}.\\  (2)  The conformal
transformation (\ref{metric}) 
  is insufficient to get an
affine-parameterized geodesic equation.
 We also need the transformation of time parameter
(\ref{paramet}).  However, the time parameter is
important when judging the chaotic behavior of a  test
particle as was pointed out in the vacuum Bianchi IX problem
\cite{Rugh}.
It may change a criterion depending on time parameter we choose. 

In GR, however,  the motion of a free test particle is
always described by geodesic, which is still interesting to
study. We do not need any conformal transformation to get
geodesic equations as we had to do for Newtonian dynamics,
so we do not have the above  problems in GR.
Hence, the second criterion based on the curvature may work well in GR.
Here we study a connection between the curvature and the
chaotic behavior of a test particle in static axisymmetric
spacetimes. We will show that the locally unstable region
(LU region) given by  the Weyl curvature tensor  becomes an
important  tool to determine  chaotic behaviors of a test
particle.

Apart from the analysis of a local instability, the
existence of an unstable periodic orbit (UPO) and
homoclinic orbit around it are also known to be  important 
causes of chaotic behavior in the Newtonian dynamics. The
location of the UPO depends on the energy and angular
momentum of the test particle. The existence of a UPO is
determined not only by the background spacetime  but also
by the orbit elements of a particle. This indicates that  a
homoclinic tangle causes strong chaos independently of the
spacetime curvature. Several  authors have shown that
strong chaos occurs in a  perturbed Schwarzschild
spacetime  via the homoclinic tangle around the
UPO\cite{Bombelli,Moeckel}. We will also study  a
homoclinic tangle of test particle motion in  static
axisymmetric  spacetimes.

 In section 2, we will quickly review a formalism for
obtaining the eigenvalues of the Riemann or 
Weyl tensor in static axisymmetric  vacuum spacetimes. In
section 3, we numerically study a correlation between chaos
and the distribution of positive eigenvalues, and show that
a local instability measured by the Weyl curvature can be
used as a sufficient condition for chaos, at least quantitatively. 
In section 4, we
will show that the UPO and a homoclinic orbit  also play
key roles in causing strong chaos. Finally, we will give
our conclusions and some remarks in section 5.
%
%
%
%
%    22    22    [Curvature and   Chaos in G.R]
%
%

\vskip .5cm
\section{Local instability  and
eigenvalues of the Weyl tensor} 
\label{sec2}\eqnum{0}
In GR, the orbit of a test
particle is described by a geodesic and it is plausible to 
examine first whether or not chaos can be predicted by the sign
distribution of a background curvature.

 The equation of a geodesic deviation $n^\mu$ is given by
the Riemann curvature tensor $R^\mu_{\: \nu\rho\sigma}$
as 
\beq
 \frac{D^2n^\mu}{D\tau^{2}}= -R^\mu_{\: \nu\rho\sigma}
u^\nu n^\rho  u^\sigma  , 
\label{deviat}
\eeq
where $u^\mu$ is the 4-velocity of a test particle
and $\tau$ is its proper time. Here we use that $n^\mu$ which
is perpendicular to  $u^\mu$, i.e., $n_\mu u^\mu =0$. 
This condition will be held at any time if we choose the $u$ and
$n$ which satisfy the condition $n_\mu u^\mu =0$ and $(Dn_{\mu}/D\tau) u^\mu =0$
initially. From (\ref{deviat}), we find 
the evolution equation for the norm of the deviation
vector, $\| n\|$, as 
\barr
\frac{d^2}{d\tau^{2}}\| n\| ={\cal K}(u,n)\| n\| 
+{1 \over 2\| n\|} \left\| n \times {Dn \over D\tau}
\right\|^2  , 
\label{31.1}
\earr
where we have used the notation $\|V\| \equiv (|V_\mu
V^\mu|)^{1/2}$ for a 4-vector
$V^\mu$.
The sectional curvature of a two-surface spanned by
$u^\mu$ and $n^\mu$ is 
{\renewcommand{\thefootnote}
{\fnsymbol{footnote}}\footnote[2]{Here we have used the
notation ${\cal R}(u,v,m,n)  \equiv R_{\:
\mu\nu\rho\sigma} u^\mu v^\nu m^\rho n^\sigma$. The sign
difference with the definition of ${\cal K}$ from
(\ref{sec_curv}) ( and (50) in \cite{Szydlo3})
 comes from the Lorentzian signature of spacetime-manifold
in GR. In our previous papers in the
Proceedings\cite{Sota}, we made a mistake in its sign, but
there is no change in our results.} 
\beq
{\cal K}(u,n) \equiv - {\cal R}(u,n,u,n)/\| n\| ^2,
\eeq
and 
$(n
\times Dn/D\tau)_\mu \equiv u^{\alpha} \epsilon_{\alpha \mu \rho
\sigma} n^{\rho}Dn^{\sigma}/D\tau 
$.

>From (\ref{31.1}), we see that a local
instability of geodesic is determined by the sign of the 
sectional curvature ${\cal K}(u,n)$. If
${\cal K}(u,n)$ is positive at some point in a
configuration space, the geodesic deviation
$n^\mu$ becomes exponentially unstable there.  The
``averaged" value of
${\cal K}(u,n)$  coincides with the
scalar curvature $R$\cite{Szydlo3}. This indicates
that a positive  scalar curvature may correspond to a
locally unstable geodesic.  However, if the
dimension of the manifold is greater than  2, the
condition 
$R>0$ does not mean that all 
sectional curvatures are positive. Then we cannot conclude
that  such an  orbit is unstable in any direction.
Moreover, in the case of geodesic in 4-dimensional spacetime, the 4-velocity $u^{\mu}$ is timelike at each tangent space. Since we cannot choose any pair of 4-vectors $(u, n)$ at each tangent space, the average for all directions seems to be useless.
We have to find an alternative way to
judge  the connection between chaos and curvature.

Here  we utilize the eigenvalues of 
the Riemann tensor which was proposed by Szydlowski et al\cite{Szydlo3}.
 We define a bivector ${\cal S}_A (A = 1 , \ldots , 6)$,
which corresponds to  
\beq
 S_{\mu\nu}\equiv u_{[\mu }n_{\nu ]} = (u_\mu n_\nu  -
u_\nu n_\mu)/2,  
\eeq
and describe the Riemann tensor in a bivector formalism as
${\cal R}^{A}_{\: B}
 (A, B = 1 , \ldots , 6)$\cite{MTW}. For a given point in
a configuration space,
${\cal K}({\cal S})$ has in general six critical
values. That is,
\beq
\partial {\cal K}({\cal S})/ \partial {\cal S}^A =0\ \ \
~~~~{\rm for}~~~~ \ \ A=1, \ldots , 6,
\eeq
if and only if ${\cal S}^A$ satisfies 
\beq
{\cal R}^{A}_{\: B}{\cal S}^B= \kappa {\cal S}^A,  
\eeq
There are six eigenvalues of ${\cal R}^{A}_{\:
B}$ and corresponding eigenvectors,  $\kappa$  and
$ {\cal S}^A$.  The critical value of ${\cal K}({\cal S})$
is equal to one of the eigenvalues in the direction of its
eigenvector. The sectional curvature in any direction is
composed of a 
 of the eigenvalues.
Therefore, the eigenvalues of the  Riemann tensor may
determine a locally unstable geodesic.\par 
 In the vacuum spacetime, the Riemann tensor
$R_{\mu\nu\rho\sigma}$ coincides with the Weyl tensor
$C_{\mu\nu\rho\sigma}$, which is decomposed into two $3 
\times 3$ symmetric matrices,  a diagonal `electric part'
${\cal E}$ and  an off-diagonal `magnetic part' ${\cal H}$
as (see  \cite{MTW})
\barr
{\cal C}=\left( \begin{array}{cc} {\cal E}&{\cal H}\\-{\cal
H}&{\cal E}\end{array}\right).
\label{32.1}\earr
\par
>From the form of (\ref{32.1}), six eigenvalues of the Weyl
tensor  are composed of three independent eigenvalues and
their complex conjugates. If the spacetime is static, the
six eigenvalues of ${\cal C}$ are real and degenerate into
three, because the magnetic part ${\cal H}$  vanishes and
the matrix ${\cal C}$ is symmetric. \par
In static
axisymmetry vacuum
 spacetime, we have two commuting 
Killing vectors, $\partial/\partial t$ and
$\partial/\partial \phi$. Then
the metric is described as 
\beq
ds^2=-e^{2U}dt^2+e^{-2U}[e^{2k}(d\rho^2+dz^2)+
\rho^2 d\phi^2], \label{33.3}
\eeq
where  $U$ and $k$ are  functions depending only on
 $\rho$ and $z$
{\renewcommand{\thefootnote}{\fnsymbol{footnote}}
\footnote[3]{ We use units of $G=c=1$, but we  explicitly
write $G$ or $c$ when it may help our discussion}. 
By using this coordinate,  we can describe the  3  $\times$ 3
matrix ${\cal E}$ as the following  form, 
\barr
{\cal E}=\left( \begin{array}{ccc}{\cal C}^{1}_{\: 1}&{\cal C}^{1}_{\:
2}&O\\ {\cal C}^{2}_{\: 1}&{\cal C}^{2}_{\: 2}&O\\O&O&{\cal C}^{3}_{\:
3} \end{array}\right)
= \left( \begin{array}{ccc}{\cal C}^{4}_{\: 4}&{\cal C}^{4}_{\:
5}&O\\ {\cal C}^{5}_{\: 4}&{\cal C}^{5}_{\: 5}&O\\O&O&{\cal C}^{6}_{\:
6} \end{array}\right),
\label{33.6}
\earr
where the suffices $1 , \ldots , 6$  denote the tetrad
components ($\hat{t}\hat{\rho}$), ($\hat{t}\hat{z}$), ($\hat{t}\hat{\phi}$), ($\hat{z}\hat{\phi}$), ($\hat{\phi}\hat{\rho}$), and  ($\hat{\rho}\hat{z}$),
respectively.  The three eigenvalues of
${\cal E}^{a}_{\: b}$ and those eigenvectors,
$\kappa_i~(i=1 , \ldots , 3)$ and 
$S^a_{\: (i)}$, are defined by
\beq
{\cal E}^{a}_{\: b} S^b_{\: (i)} = \kappa_i  S^a_{\:
(i)}~~~{\rm for} ~~~i=1, \ldots , 3 \eeq
>From the form (\ref{33.6}),  the
component ${\cal C}^{3}_{\: 3}={\cal C}^{6}_{\: 6}$ is one
of the eigenvalues, 
$\kappa^c_{03}$ .  We also find the remaining   eigenvalues as
\barr
\kappa^c_{01} & = & \frac{1}{2} \left[ {\cal C}^1_{\:
1} + {\cal C}^2_{\: 2} + \sqrt{({\cal C}^1_{\: 1}-{\cal
C}^2_{\:2})^2+4({\cal C}^1_{\: 2})^2}\right] ,\non\\
\kappa^c_{02} & = & \frac{1}{2}\left[{\cal C}^1_{\:
1} + {\cal C}^2_{\: 2} -  \sqrt{({\cal C}^1_{\: 1}-{\cal
C}^2_{\:2})^2+4({\cal C}^1_{\: 2})^2}\right].
\label{koyuu2}
\earr
Since the trace of the Weyl tensor  (and then the trace of
the electric part ${\cal E}$) vanishes,  the   sum of the
eigenvalues
$\kappa^c_{0i}$ also vanishes, i.e., 
\beq
\kappa^c_{01}+\kappa^c_{02}+\kappa^c_{03}=0.
\label{wa}
\eeq
 In this case, we can divide the spacetime first into two
regions by the signature of the  eigenvalues. One is a
region where two eigenvalues are positive and the rest  is
negative,  and  the other is a region with two negative and
one positive  eigenvalues.   We shall further classify  the
former case into two types, i.e., one is the region where
$\kappa^c_{01}>0, \kappa^c_{02}<0$ and $\kappa^c_{03}>0$
($[+,-,+]$-region)  and the other is the region where
$\kappa^c_{01}>\kappa^c_{02}>0$ and
$\kappa^c_{03}<0$ ($[+,+,-]$-region). 
 We shall call the $[+,+,-]$-region a locally unstable
(LU) region, which becomes important for chaos.  The
reason is as follows: First, since positive eigenvalues
contribute to a local instability, the cases with two
positive eigenvalues are likely to be more unstable
locally than the case with just one positive eigenvalue. 
Although there are two regions with two positive
eigenvalues (the $[+,-,+]$-, $[+,+,-]$-regions), the latter one
(LU region), in which $\rho$-$z$ plane orthogonal to
 the two Killing directions becomes unstable
seems to be important because
an instability in a Killing direction
may not play any role in chaotic behavior of the orbit.
 Secondly, the
LU region does not appear in the Petrov type D
spacetime, in which a test particle motion is 
integrable\cite{Kramer} and does not show any chaotic
behavior, but the other type of region exists even in the
type D solutions.  Hence we expect that the LU region
 seems to be more important for a local instability
 than any other region.  It will be confirmed by numerical
analysis in the next section. In the next section, we will numerically
study a test particle motion in several background
spacetimes and show a close relation between the
above-defined LU region and chaotic behaviors of test
particles. %
%
%
%
%3    333\par3    333\par3    333\par3    333\par3  
%   
%
%

\vskip .5cm
\section{Numerical analysis of local instability}
\label{sec3}\eqnum{0} 
In this section  we study a  correlation between the LU
region and  chaos of a test particle. We will show that the
LU region given by the  Weyl tensor plays a key role in
causing the chaotic behavior of test particles. We analyze
the chaotic behavior of a test particle by using the
Poincar\'e map and the Lyapunov exponent. To define the
Lyapunov exponent in GR, we use the proper time of each
geodesic as a time-flow parameter. (see also \cite{Karas,Cornish}
for another choice.) We also define a measure distance 
$\Delta$ in a  phase space as
\beq
\Delta^2=g_{\mu\nu}n^\mu n^\nu
+g_{\mu\nu}\frac{Dn^\mu}{D\tau}\frac{Dn^\nu}{D\tau}.
\label{measure}
\eeq
The proper time, $\tau$, should be normalized by a typical
time scale in the system, e.g.,  $GM/c^3$, where $M$  is a
mass of a localized object. We can easily show that 
(\ref{measure}) is  positive definite along each geodesic,
because  both
$n^\mu$ and $Dn^\mu/D\tau$ are spacelike.
So in our case we do not face the
problem which Biesiada et.al. pointed out when defining a
measure distance in a phase space with a Lorentzian
signature
\cite{Biesiada}.  By using this  measure, we   calculate the
maximum Lyapunov exponent, which is defined by 
\beq
\lambda
=\lim_{N\rightarrow\infty}\frac{1}{N\delta
\tau}\sum_{n=1}^{N}\ln [\Delta  (n\delta \tau)], \label{Lyp}
\eeq
where $\Delta (n\delta \tau)$ is the value of $\Delta$ 
after
$n\delta \tau$ time evolution, which is normalized to a
unit length after each time interval $\delta \tau$
\cite{Lypnov}. The reciprocal of $\lambda$ can be utilized
as a standard time scale where the effect of chaos becomes
conspicuous. We have integrated both the geodesic
 and the deviation equations using the predictor-corrector
and the Burlisch-Store integrators \cite{recip}. We have
not found any remarkable difference between  the two methods.
\par

We also map the   LU region in  static  axisymmetric vacuum
spacetimes and examine  the chaotic behavior of test
particles passing through such a region  by  the Poincar\'e
map. The allowed region where a test particle  can move
is obtained from the super-Hamiltonian 
$$ H=\frac{1}{2}g_{\mu\nu}\left(\frac{d{x}^{\mu}}{d\tau}\right)
\left(\frac{d{x}^{\nu}}{d\tau}\right).$$
Since $H$ is conserved and fixed on the value $-1.0$, we find that the 
square of meridian velocity 
\barr
v_{\ast}^2 &\equiv & \frac{1}{2}\left( g_{11}\left(\frac{d{x}^{1}}{d\tau}\right)^2+g_{22}\left(\frac{d{x}^{2}}{d\tau}\right)^2\right)\non\\
&=&\frac{1}{2}\left(\left(\frac{d\rho}{d\tau}\right)^2+\left(\frac{dz}{d\tau}\right)^2\right)\exp{(-2U+2k)}
\label{vhosi2}
\earr
 for the particle with energy $E$ and angular momentum $L$ can be expressed as
\beq
v_{\ast}^2(\mbox{\boldmath $x$}, E, L)=
\frac{(E^2-V^2_{\rm eff}(\mbox{\boldmath $x$}, L))}
{2\|\partial/\partial t \|^{2}},
\label{vhosi}
\eeq
where 
$\mbox{\boldmath $x$}=(\rho, z)$ 
and $\|\partial/\partial t \|^2 \equiv -g_{00}$.
$V^2_{\rm eff}$ is
 the effective potential
 for the particle with the angular momentum
$L$, which 
  is defined by using the norms of two Killing
vectors,  $\partial/\partial t$ and
$\partial/\partial \phi$ as
follows. 
\beq
V^2_{\rm eff} (\mbox{\boldmath $x$}, L) \equiv \|\partial/\partial t \|^2 
 \left(1+\frac{L^2}{\|
\partial/\partial \phi \|^2 }\right),
\eeq
where $\| \partial/\partial \phi \|^2 \equiv g_{3 3}$.
Since $v_{\ast}^2$ and $\|\partial/\partial t \|^2$ are positive in (\ref{vhosi}), the allowed region $D_{\rm eff} $ for the particle with $E$ and $L$ is given  as
\beq
D_{\rm eff} \equiv \{\mbox{\boldmath $x$}|V^2_{\rm eff} (\mbox{\boldmath $x$}, L) \leq E^2 \}. 
\label{veffe}
\eeq
If $D_{\rm eff} $ is compact, i.e., if a test particle 
is trapped and never falls into a
singularity, we call the particle geodesic  a bound orbit.
We discuss a chaotic behavior only for this bound orbit.

\subsection{Numerical results}
The methods to find exact
solutions  in  a static axisymmetric
vacuum case   are well-developed
\cite{Zip}--\cite{Azuma}. We can easily construct exact
solutions with any mass distribution.  In this paper, we
use several known exact solutions to study a test particle
motion.

First we shall analyze the Zipoy-Voorhees (ZV) solution
\cite{Zip}. The functions in a spacetime metric (\ref{33.3}) are
\barr
U &=&\frac{\delta }{2} 
\ln\left(\frac{r_1+r_2-2m}{r_1+r_2+2m}\right) \label{21.5}\\
k&=&\frac{\delta^2}{2}\ln\left[\frac{(r_1+r_2+2m)(r_1+r_2-2m)}{4r_1
r_2}\right], 
\earr
where $r_1=[\rho^2+(z-m)^2]^{1/2}$,
$r_2=[\rho^2+(z+m)^2]^{1/2}$ and $m$ is a mass parameter.
The  singularity exists between $\pm M/\delta$ on z axis.
Here $M$ is the gravitational mass of the singularity which
is given as $m\delta$ by using the parameters $m$ and
$\delta$. The quadrupole moment $Q$ is also given as
$m^3\delta^3(1-1/\delta^2)/3$.  In the case of $\delta =1$,
$Q$ vanishes and the spacetime just becomes 
the  Schwarzschild solution. For the
 Schwarzschild spacetime, no LU region appears because
 two eigenvalues of ${\cal E}$ are degenerate and 
negative. This comes from the fact that the Schwarzschild
spacetime belongs to the Petrov type D.  In the
limit of $\delta \rightarrow \infty$ fixing $M=m\delta$,
the singularity shrinks to the origin and the
spacetime coincides with the Curzon spacetime\cite{Curzon},
with metric functions, $U$ and $k$, given by 
\barr
U&=&-G\frac{M}{r} , \non\\
k&=&
-\frac{1}{2}G^2\frac{M^2\rho^2}{r^4},
\label{cul}
\earr
where $r\equiv (\rho^2+z^2)^{1/2}$.
For the case  of $\delta \neq 1$, the eigenvalues 
of ${\cal E}$ are not degenerate, but both of the
eigenvalues,
$\kappa^c_{01}$ and $\kappa^c_{02}$ are not positive and no LU
region appears (Fig.1). We have  found that for any bound orbit
both in the ZV spacetime and in the Curzon spacetime, the
sectional curvature is always negative and  no chaotic
behavior of the orbit is seen at least  from our analysis
by  use of the Lyapunov exponent and the Poincar\'e map.

Secondly we examine  a system
with $N$-point Curzon-type singularities on the symmetric
axis \cite{Ncul}, whose metric functions,
$U$ and
$k$, in (\ref{33.3}) are 
\barr
U&=&-G\sum_{i=1}^{N}\frac{M_{i}}{r_{i}} , \non\\
k&=&-\frac{1}{2}G^2\sum_{i\neq j}^{N}\frac{M_{i}
M_{j}}{r_{i}r_{j}}\left[1-\left(\frac{r_{i}-r_{j}}
{b_{i}-b_{j}}\right)^2\right]
-\frac{1}{2}G^2\sum_{i=1}^{N}\frac{M_{i}^2\rho^2}{r_i^4}
\label{ncul},
\earr
where  $M_{i}$ and $b_{i}$ denote the mass and  the
position parameters of $i$-th singularity on z axis,
respectively, and
$r_i\equiv [\rho^2+(z-b_i)^2]^{1/2}$ \footnote[1]{In 
\cite{Ncul}, the  last term in (\ref{ncul}) seems to be
missed, otherwise  the vacuum Einstein's equations are not
satisfied.}. We call this solution the 
$N$-Curzon spacetime because the $N=1$ case corresponds to  the Curzon
solution.

\par We have  first analyzed 
the case with
 a reflection symmetry on the equatorial plane.  For the
N-Curzon  spacetime, 
LU regions exist and some bound orbits of test
particles  intersect within them. As we see from Fig.2, in
2-Curzon spacetime, the LU region appears on
the equatorial plane. If the energy of the orbit gets
larger than that at the UPOs ($E_{\rm UPO}$), it fails to
be bounded, so we  just increase the energy up to $E_{\rm
UPO}$ for a given angular momentum $L$. 
When  the energy approaches to $E_{\rm UPO}$ and
 it eventually exceeds some critical value $E_{\rm cr}
(<E_{\rm UPO})$, tori in the phase space are broken (Fig.3).
  and strong chaos occurs in those orbits. 
On the other hand, the orbit which
does not pass the LU region does not show any chaotic
behavior, even if the orbit has the same energy as the 
chaotic one (Fig.4).

In 3-Curzon spacetime, where one of
the point-mass singularities is at the origin and the other
two are on the z axis at the same distance from the origin,
two LU regions appear between those singularities. The
result is similar to the case of 2-Curzon spacetime.  Some
orbit departs from the equatorial plane and  eventually
approaches to the LU region. Then, just after it crosses
the LU region, the torus begins to break and the orbit
becomes strongly chaotic (Fig.5).  On the other hand, no
bound orbit not passing through the LU region shows any
chaotic behavior(Fig.6).

Next we consider the spacetime with $N$ ZV-type singularities
put on z axis in order to see how the shape of singularity effects
on the chaotic property of geodesics.
It is known that a solution of $N$ ZV-type 
singularities located  on the z axis is derived by the
inverse scattering method, where $N$ is an arbitrary
natural number
\cite{Azuma}. The most general form of the $N$-soliton
solution  with real-poles is given by 
\barr
U&=&\frac{1}{2}\ln \left(i^N \frac{\zeta_1\zeta_2
....\zeta_N} {\rho^N}\right)^\delta\non\\
k&=&\frac{1}{2}\ln \left[\frac{\rho ^{N^2/2}
{\displaystyle\prod_{n>l}}
(\zeta_n-\zeta_l)^2} {{\displaystyle \prod_n}
(\rho^2+\zeta_n^2) {\displaystyle
\prod_l}\zeta^{N-2}_l C^{(N)}}\right]^{\delta^2} ,  \label{azu1}
\earr
where
\barr
\zeta_n&=&w_n-z\pm\sqrt{(w_n-z)^2+\rho^2}\non\\
C^{(N)}&=&2^{\frac{N}{2}(N-2)}
\prod_{n>l}^{N/2}
(w_{2n-1}-w_{2l-1})^2(w_{2n}-w_{2l})^2. \label{azu2}
\earr
In the case $N$=2, if $w_1$ and $w_2$ are
parameterized as $-m$ and $m$ and the signs in $\zeta_1$
and $\zeta_2$ are chosen to $+$ and $-$, respectively, the
solution (\ref{azu1}) is asymptotically flat and coincides
with the  ZV solution. When $N$=4, setting $w_1=z_1-m_1$,
$w_2=z_1+m_1$, $w_3=z_2-m_2$ and
$w_4=z_2+m_2$, and choosing the signs in $\zeta_1 \sim
\zeta_4$ as
$(+,\ -,\ +,\ -)$, the solution (\ref{azu1}) is
asymptotically flat and corresponds to  the spacetime which
contains two ZV type singularities with  masses  
$m_1\delta$  at $z_1$ and  $m_2\delta$  at $z_2$ on z axis.
We call this   2-ZV solution. 

Here we examined  2-ZV solution with  two singularities at $\pm 2 GM/c^2$ on the z axis in the case of $\delta= 1.0$. In this case, LU region appeared on equatorial plane in the same way as 2-Curzon spacetime and some of the bound orbits  passing through the LU region become strongly chaotic (see Fig.7). 

>From the above analysis, the appearance of the LU region seems
to be closely related to the chaotic behavior of
geodesics. We may conclude that passing through the LU region is
necessary to cause the chaotic motion of geodesic and the upper bound of $\lambda$ is determined by the value of positive eigenvalue $\kappa_1$ and $\kappa_2$ in an LU region.

\subsection{Lyapunov exponent and Curvature }
So far we have shown that the LU region plays a crucial role in causing strong chaos in a static axisymmetric spacetime, qualitatively. However, in order to apply chaos to real astrophysical phenomena, it is necessary to extract  some quantitative information about the strength of chaos from the LU region.  One way is to estimate a Lyapunov exponent $\lambda$. We show our results in Fig.8 for 2-Curzon spacetime and Table 1 for 2-ZV spacetime.  In Fig. 8, we choose the same situation and initial condition for the orbit as in Fig.2, 3, and find the result that the inverses of Lyapunov exponents ($\lambda^{-1}$) are about $33 \tau_P$ and $4.7
\tau_P$ for (b) and (c), respectively, where $\tau_P$ is the average period around the symmetric axis for each particle orbit. For the 2-ZV spacetime , we studied two additional values $\delta$ ($\delta=0.65, 10.0$) as well as the case of $\delta=1.0$ depicted in Fig.7, and 
found that
 as $\delta$ increases, the value of $\lambda$ becomes smaller and smaller (see Table 1).
In this case, for each $\delta$,
 $\lambda^{-1}$ becomes (a)$10.8 \tau_P$, (b)$9.4 \tau_P$ and (c)$8.0 \tau_P$, respectively.
These results indicate that the timescale on which chaos becomes conspicuous is about several  Keplerian rotations. So the chaos we found may be strong enough to be potentially observed around compact objects in which the effects of GR are important.

As is shown in the above, a local instability is closely related to the occurrence of chaos, giving us the prospect of evaluating the Lyapunov exponent $\lambda$ in terms of the curvature. 
Here we will examine it from a quantitative viewpoint. 
Since the local instability  is determined by the sectional curvature
${\cal K}(\mbox{\boldmath $x$},u,n)$ as we can see from (\ref{31.1}), $\lambda$ for chaos around LU region may be determined by some average of ${\cal K}(\mbox{\boldmath $x$},u,n)$ as
\beq
\lambda \approx \sqrt{<{\cal K}>},
\label{sum3}
\eeq
where $<{\cal K}>$ is the averaged value of ${\cal K}(\mbox{\boldmath $x$},u,n)$ inside the bound region $D_{\rm eff} $ which overlaps an LU region. In vacuum spacetime, the sectional curvature ${\cal K}(u,n)$ at each point $\mbox{\boldmath $x$}$ for given four velocity $u=u^{(\alpha)}e_{(\alpha)}$ and deviation vector $n=n^{(\alpha)}e_{(\alpha)}$, can be expressed
 as a linear combination of $\kappa^c_{0i}$, 
as follows
\barr
{\cal K}(\mbox{\boldmath $x$},u,n)=\sum_{i=1}^{3} A_{i}(u,n)\kappa^c_{0i}
(\mbox{\boldmath $x$}),
\label{sum}
\earr
where  $$A_{i}(u,n)\equiv [(u^{(0)}n^{(i)}-u^{(i)}n^{(0)})^2
-(u^{(j)}n^{(k)}-u^{(k)}n^{(j)})^2]/\| n\|^2,$$
($(i,j,k)$ is a permutation of $(1,2,3)$).
Since we can eliminate $\kappa^c_{03}$  by the condition (\ref{wa}),
the value of $<{\cal K}>$ is determined by some
average of $\kappa_{1}$ and $\kappa_{2}$.
Here we utilize the geometrical mean, defined as
\beq
<\kappa>\equiv \sqrt{\kappa_{1}\kappa_{2}(\mbox{\boldmath $x$})},
\label{heikin2}
\eeq
since $<\kappa>$ vanishes at the boundary of the LU region and takes the maximum value in the middle of the region (see Fig.6). 
 Estimating $\lambda$ from $<\kappa>$ as defined in (\ref{heikin2}) seems to
be a reasonable rough estimate, since other averages, such as the arithmetic mean of $\kappa_{1}$ and $\kappa_{2}$ give the same order of magnitude as $<\kappa>$ at least in the vicinity of LU region.
We estimate the average values of $<\kappa>$ in the intersection of the 
LU region and $D_{\rm eff}$ as $\sqrt{<\kappa>_{max}}$, where
$<\kappa>_{max}$ is the maximum value of $<\kappa>$ in the intersection
of the two regions. In Table1, we compare $\sqrt{<\kappa>_{max}}$ with the
actual value of $\lambda$ calculated for several chaotic orbits passing
through the LU region.
Unfortunately, the value of $\lambda$ differs by approximately one from an averaged value $\sqrt{<\kappa>_{max}}$ for each
$\delta$ in Table1.
This deficit comes from the fact that the chaotic orbit 
not only moves on the meridian plane, but also rotates around the z axis. This rotation restricts the movement of the geodesic on the meridian plane according to the condition, $v_{\ast}^2\geq 0$ in (\ref{vhosi}).
In fact, we do not see any chaotic motion for a geodesic
passing through an LU region, if we choose the $E$ and $L$ so as to make the geodesic a stable periodic orbit. We also showed in 
Fig.3 that for the fixed value of $L$, the chaos becomes 
stronger and stronger around an LU region as $E$ increases and approaches the
value for the UPO. From (\ref{vhosi}), these results mean that chaos becomes more and more conspicuous as the meridian velocity $v_{\ast}$ increases.
So it is natural to expect that $\lambda$ can also be affected  by $v_{\ast}$. Since each component $A_{i}(u,n)$ in (\ref{sum}) is proportional to the combination of two components of $u$, we utilize $v_{\ast}^2$ and compare the product of its averaged value and $<\kappa>$ with the numerical value of $\lambda$.
Here we take the maximum value of $v_{\ast}$, i.e., $v_{\ast,max}$,
as the average value of $v_{\ast}$ inside the part of the LU region that overlaps $D_{\rm eff}$.

As shown from Table 1, our estimate agrees well with the real Lyapunov exponent $\lambda$ for the most chaotic orbit for each value of $\delta$. As far as asymptotically flat vacuum spacetimes are concerned, the bound condition (\ref{veffe}) restricts the value of $v_{\ast,max}$ to that less than the velocity of light. It reduces the estimate of $\lambda$ for each geodesic  to almost one tenth of $\sqrt{<\kappa>_{max}}$, which is determined just from the curvature itself. It seems that $\sqrt{<\kappa>_{max}}$ gives at most the upper limit of the Lyapunov exponent for the relativistic orbit passing through an LU region.
The $v_{\ast}/c$ dependence in $\lambda$ is also important when characterizing the chaotic motion of geodesic in GR, in contrast to free particle motion in Newtonian mechanics, because the velocity dependence of local instability in GR completely disappears  in the Newtonian limit, as we will show in the next
subsection. So although our quantitative estimate of $\lambda$ is rather empirical, the $v_{\ast}$ dependence of sectional curvature ${\cal K}(u,n)$ seems to be essential, in order to make the criterion of local instability work well in GR.
\subsection{Newtonian Dynamics vs. General Relativity}
The results in
\S. 3.1  show that the LU region  plays a crucial role in
causing strong chaos at least in the vacuum spacetime, but
using similar criteria to assess the presence of chaos by
analyzing local instability in Newtonian dynamics is not
always successful, as mentioned in \S 1.  Here we shall
compare our criterion in GR and those in Newtonian theory
and show the reason why our criterion does work. 
\par 
There are two criteria for local instability in Newtonian
theory as discussed in \S 1: One is by the eigenvalues of
the tidal matrix, and the other is the curvature of a
fictitious space obtained by a conformal transformation. In
\S  1, we pointed out several  obstacles in the latter 
approach.  Although our approach is technically quite
similar to this,  we do not face such obstacles, because we
have a geodesic equation without conformal transformation
and study test particle motion in a physical spacetime. 
The farmer approach in Newtonian theory is physically much
closer to ours in GR, as we will discuss below.
 
To show the relation between a criterion in the Newtonian
theory and that in GR, we take the Newtonian limit  of the
deviation equation (\ref{deviat}). If the spatial
components of the 4-velocity $v^i \equiv u^i/u^0$ are much
less than that of light $c$, we can approximate the
4-velocity as $ u^\mu
\approx (c,\ v^i)$. The deviation equation (\ref{deviat})
becomes
\beq
\frac{D^2 n^i}{D\tau^2}=-c^2R^i_{\: 0j0}n^j-cR^i_{\: 0jk}n^jv^k 
-cR^i_{\: kj0}v^kn^j-R^i_{\: jkl}n^kv^j v^l \label{devkinji}\ .
\eeq
We also assume that the spacetime is approximately a flat
Minkowski spacetime with small perturbations, i.e.,  the
metric is expressed as $g_{ab}=\eta _{ab}+h_{ab}\ (h_{ab}
\ll 1)$, where $\eta _{ab}$ denotes the Minkowski metric.
The Riemann tensor $R_{\alpha \beta
\gamma \delta }$  is now
\barr
R_{\alpha \beta \gamma \delta }=\frac{1}{2}(
\partial_\beta\partial_\gamma h_{\alpha\delta}
+\partial_\alpha\partial_\delta h_{\beta\gamma}
-\partial_\alpha\partial_\gamma h_{\beta\delta}
-\partial_\beta\partial_\delta h_{\alpha \gamma}). 
\label{riekinji}
\earr
Introducing the Newtonian potential $V$ as $V\equiv
-c^2h_{00}/2$, $R^i_{\: 0j0}$ becomes 
$\partial^i\partial_j V /c^2$. To the leading order of
$h_{ab}$ and $v^i/c$, the terms with
 $v^i$ in (\ref{devkinji}) do not appear and
Eq.(\ref{devkinji}) coincides with the tidal acceleration
equation (\ref{tidal}) in the Newtonian theory. To the same
order, the sectional curvature (\ref{sum}) becomes 
\beq
{\cal K}(\mbox{\boldmath $x$},u,n)=
-[(n^{1})^2k_{1} +(n^{2})^2k_{2}
+(n^{3})^2k_{3}]/c^2,  \label{kappkin} 
\eeq 
where
$k_{i} (i = 1 \sim 3)$ are the eigenvalues of
$V^i_{~j}$ and correspond to
$-c^2\kappa^c_{0i}$\footnote[2]{The sign difference comes
from the fact that we treat the Euclidean space in the
Newtonian theory, while the Lorentzian spacetime in GR.}
\par  From these equations, 
we find  that 
the deviation equation in the Newtonian limit becomes 
adiabatic and lose any  information of the velocity of the
orbit.  For an axisymmetric system in the Newtonian theory,
we can also define an LU region  as we did in GR, i.e.,  
a region where both $k_{1}$ and $k_{2}$ are negative. Then,
for  deviations perpendicular to the symmetric direction of
$\partial/\partial
\phi$, all orbits become locally unstable in an LU region regardless of those velocities  because $n^{3}=0$ in  (\ref{kappkin}) for such deviations.
It seems that the same criterion for chaos applies in the
Newtonian case. 
\par 
However, this is not true.  For
example, we can find an LU region in the Newtonian model
where two gravitational point sources are fixed on z-axis.
Even if some bound  orbit passes through the LU region, (in
fact, we can find such an orbit), no chaotic behavior
should be found  because of the integrability of the
system\cite{Charl}. The reason why the  LU region does not
predict strong chaos is that its adiabatic nature cuts  the
connection between local instability described by an LU
region and the global instability of a bound orbit. This is confirmed
from our previous analysis in Sec.3.2, in which the Lyapunov
exponent $\lambda$ depends on a particle velocity. Our result is also
consistent with the failure of the use of the eigenvalues
of the tidal matrix as a criterion for chaos in the
Newtonian theory.

%
%
%
%4    444\par4 444\par4 444\par4 444\par4 
%   
%
%

\vskip .5cm
\section{Homoclinic tangle around UPO and chaos in GR}
\label{sec4}\eqnum{0} 
In the previous section, we have shown that the LU region
plays an important role in causing chaotic behavior of a
test particle. However,  in 2-Curzon spacetime, we have
also found some  orbits whose chaotic behavior cannot be
explained by the passage through the LU region. In Fig.8,
where two singularities with different mass  ($M$ and
$0.5M$) are put at  $z=-4M$ and $4M$ on the z axis, we find
chaotic behavior for test particles which do not pass
through the LU region. In 2-ZV spacetime, we have found the
similar examples. Hence, we seem to have another type of
chaos, which is not understood by the criterion discussed
in Sec.3.

Here we will explain such exceptional cases from another
point of view, that is, the existence of UPO and a 
homoclinic tangle
around it.  In GR, a single point-mass system such as the 
Schwarzschild spacetime usually has a UPO. Around the UPO
in the  Schwarzschild spacetime, there exists an orbit  of
a test particle which gradually departs from the UPO point
and eventually returns to it with infinite time interval,
if the angular momentum is  appropriately chosen. This
 orbit is called a homoclinic orbit  and is known to play
an important role in causing strong chaos. A heteroclinic
orbit may also be defined as the orbit which gradually
departs from a UPO point, but approaches another UPO point
with infinite time interval.

For spacetimes with a reflection symmetry on the equatorial
 plane, we find a
homoclinic orbit which departs from and returns to the UPO 
on the plane. This orbit is not chaotic. 
However, if some perturbations by other gravitational
sources break    such a reflection symmetry, the homoclinic
(or heteroclinic) orbit  no longer exists, and stable and
unstable manifolds starting with the UPO point could be 
split and tangled in a complicated way. This splitting  of
the homoclinic orbit is called a homoclinic tangle and is
known to cause strong chaos \cite{Homo}.

For example, Bombelli and Calzetta
\cite{Bombelli} showed by using the Melnikov method that 
when the Schwarzschild spacetime is perturbed by linear
gravitational waves, strong chaos  in  a test particle
motion appears through the tangle of a homoclinic orbit
around a UPO point. Moeckel also  showed by the similar 
method that strong chaos  appears   in the Schwarzschild
spacetime with  different perturbations,  whose Newtonian
limit reduces to the famous Hill's problem\cite{Moeckel}. 

In our case, if the background spacetime has a reflection
symmetry as discussed in the previous section, there exists
a homoclinic orbit on the equatorial plane.  However, for
the spacetime without a reflection symmetry, a homoclinic
orbit does not exist and stable and unstable manifolds
starting with the UPO point are tangled.  Such a tangle is
realized by breaking a reflection symmetry. In fact, as we
have shown in Fig.8,  the orbits become strongly chaotic 
even if they do not pass through any LU region. 

 We can easily distinguish this  type
of chaos from the previous one studied by a local
instability. Choosing the equatorial plane as the
Poincar\'e section, we find
that   a homoclinic orbit, if it exists, will appear on the
boundary of a family of tori. Then we can easily check the
occurrence of a homoclinic tangle by examining whether or
not the  boundary of tori is broken. For the case with a
reflection symmetry, the boundary torus in the Poincar\'e
map is not broken as seen in Fig.2, while 
 in Fig.8, we  see the boundary  torus is broken, which
means that a homoclinic tangle occurs in the case without a
reflection symmetry.
\par
 In general, a homoclinic tangle requires periodic
perturbations around a homoclinic orbit, because a stable
manifold tangles with an unstable one infinitely  through a
periodicity of those manifolds. However our case does not
need a periodicity of   perturbations, because a homoclinic
orbit itself is periodic around the z axis even before  it
is perturbed because of the axisymmetry. So it is enough to
break the reflection symmetry on the z axis in order to
cause an infinite tangle of stable manifold with unstable
one, as we  find  in our numerical results.

 A homoclinic tangle
strongly depends on the properties of each geodesic such
as the energy or angular momentum. So it is plausible that
in this case the background curvature, which is independent
of the properties
 of each geodesic,
no longer determines the chaotic behaviors of a test
particle. This may be the reason why we have seen 
strong chaos outside the LU region.
%
%    5555    5555    5555    5555   5555
%   
%    Concluding Remarks
%
%\vspace{7cm}

\vskip .5cm
\section{Concluding Remarks}\label{sec5}\eqnum{0} 
We have examined two criteria for the chaotic motion of a
test particle in static axisymmetric spacetime. We find
that unlike  Newtonian dynamics, there is a close relation 
in GR between chaos and the curvature of background
spacetimes. We have used the
eigenvalues of the Weyl tensor instead of the  scalar
curvature as the criterion for chaos, because it can be
applied even in the vacuum spacetime. In a
point-singularity system, the LU region does not appear and
no chaotic behavior is seen regardless of the ``shape" of
the singularity. On the other hand, in the
plural-singularity systems such as
$N$-Curzon spacetimes or 2-ZV spacetime, the LU regions
appear between those singularities and we find that some 
orbits become strongly chaotic there, if the energy
increases beyond some  critical value. 
\par
The Petrov type D spacetime is integrable and  the LU
region does not exist for such a spacetime. From our
analysis, we can conclude that the
 existence
of the LU region  may be  closely related to the
non-integrability of the spacetime as well as  chaotic
behavior of a test particle.

However we know that only the passing through this region does not determine the fate of the orbit completely. As we showed in Fig.2, the behavior of an orbit crossing an LU region depends strongly on its energy $E$ and angular momentum $L$. If the energy $E$ is less than some critical value, the orbit does not become chaotic even in an LU region. This is because the orbit does not move on the meridian plane orthogonal to the two Killing directions, while the LU region causes an instability in the direction parallel to this plane. In fact, a stable circular orbit can exist in an LU region for appropriate values of $E$ and $L$. It is never chaotic because it is strictly restricted to move on the meridian plane. This can be explained by the fact that a meridian motion determined by $v_{\ast}$ vanishes from (\ref{vhosi}) for a stable circular orbit.
 It seems that the LU region can well cause chaotic motion only when the orbit has a high enough meridian velocity, $v_{\ast}$ to be able to move freely on meridian plane. Our numerical results in sec.3.2 give empirical support
for the $v_{\ast}$ dependence of $\lambda$.

As we showed in Sec.3.3, the $v_{\ast}$ dependence of $\lambda$ is also important when comparing chaotic motion of geodesics in GR to free particle motion in Newtonian mechanics. In the Newtonian case, the $v_{\ast}$ dependence on the sectional curvature disappears and local instability cannot be used to estimate chaos. So $v_{\ast}$ seems to play a key role in making LU region
function well as the criterion for chaos in GR.

Although our criterion is a sensitive test for the occurrence
of chaos, it is no more than a sufficient condition. In fact,
we have also shown that the existence of the  UPO and the
homoclinic tangle around the UPO is an important cause of
strong chaos. In particular,  when a homoclinic tangle
appears through  perturbations, strong chaos occurs around
a UPO even if the orbit does not pass through the LU region.
In order to make the criterion more reliable, it seems to be necessary to include information about each geodesic, such as $E$ and $L$.
The work of Szydlowsky et.al. develops this idea. They devised a way to judge local instability by utilizing the eigenvalues of the Riemann tensor for the geodesic after the conformal transformation, including $E$ \cite{Szydlo}. However, this kind of approach that uses a conformal transformation has difficulties, as we pointed out in introduction. 
In fact,  U. Yurtsever\cite{Yurt} showed that the curvature
determined after the conformal transformation does not always
accurately predict the presence of chaos in static spacetime.
So the problem will have to be considered further in the future. 
 
>From the astrophysical point of view, it is likely  that
such a homoclinic tangle could occur around  compact
objects with UPOs through perturbations by gravitational
waves or other gravitational sources such as stars or
galaxies. However,  our results indicate that 
we have to take into
account the curvature effect, which is characterized
by LU region in axisymmetric static spacetimes,
 independently as a source of
strong chaos.  The reciprocal of the maximum Lyapunov
exponent we got in these kinds of chaos  is about  several
rotations of  the Keplerian orbit. So the chaos characterized
by curvature 
may also be expected to be observed around the compact objects
in which the effect of GR is important. The general
relativistic chaos determined by both of the two causes
could be necessary to
explain some unknown relativistic  phenomena in
astrophysics in the future. 

In this paper we restrict our analysis to  the case of
static axisymmetric  vacuum spacetimes. In this case, the
Riemann tensor coincides with the Weyl tensor and  the six
eigenvalues degenerate into three, which makes our
analysis  simple. It is natural to ask whether or not our
results hold for more general cases. If spacetime is not
empty,  the Riemann tensor is not the same as the Weyl
tensor by the effect of local matter fluid and   six
eigenvalues become independent. Although the curvature
analysis become more complicated, the way to analyze it is
straightforward. We will present our analysis for the
static case with this matter effect  and show that our
present results are still valid in the next paper. For
stationary spacetimes with rotation, however, it turns out
to  be  much more difficult to analyze, because  the
curvature ${\cal R} $  is not  symmetric and  its
eigenvalues  become complex. We are also very interested in
non-stationary system such as a coalescing binary system, where
we cannot introduce an effective potential. In spite of
those difficulties, we believe that a sectional curvature
and a homoclinic tangle are closely related to strong
chaos.  Such an  analysis will be undertaken  in the
future. 

\vskip 1cm

\noindent
-- Acknowledgments --
We would like to thank H. Yoshida and T. Harayama
 for useful discussions and information. We also
acknowledge R. Easther for his critical reading of our
paper. This work was supported partially by the
Grant-in-Aid for Scientific  Research Fund of the Ministry
of Education, Science and Culture   (No. 06302021 and No.
06640412) and by the Waseda University Grant  for Special
Research Projects. 
% %
%
%    bbbb    bbbb    bbbb    bbbb    bbbb
%

\newpage
\begin{table}[hbpt]
%Each $\lambda$ corresponds to the Lyapunov exponent of the chaotic %orbit in Fig.7.}
\begin{center}
\renewcommand{\arraystretch}{2.0}
\begin{tabular}{|c||c|c|c|c||c|c|}
 \hline
case & $\delta$ & $Q/M^3$ & $\lambda/M^{-1}$  & $v_{\ast,max} \sqrt{<\kappa>_{max}}/M^{-1}$ & $v_{\ast,max}(E, L)$
 & $\sqrt{<\kappa>_{max}}/M^{-1}$ \\
\hline  \hline
(a) & 0.65 & $-$0.4556 & 4.6537$\times10^{-3}$ & 4.3574$\times10^{-3}$ & 5.3412$\times10^{-2}$ & 8.1580$\times10^{-2}$ \\
 \hline
(b) & 1.0 & 0.0 & 4.2510$\times10^{-3}$ & 3.9369$\times10^{-3}$ & 5.6232$\times10^{-2}$ & 7.0010$\times10^{-2}$ \\
\hline
(c) &10.0 & 0.3333 &  3.7543$\times10^{-3}$ & 2.7760$\times10^{-3}$ & 4.2481$\times10^{-2}$ & 6.5347$\times10^{-2}$ \\
\hline
\end{tabular}
\label{tab5-1}
\end{center}
\caption{}
\end{table}

\begin{flushleft}
{Figure Captions}
\end{flushleft}
\baselineskip = 24pt
%%%%%%
\parbox[t]{2cm}{\bf FIG 1:}\ \ 
\parbox[t]{14cm}
{The $[+,-,+]$-region  (shaded)  of the Curzon spacetime 
with mass parameter, $M$, and the bound region (dotted)
for a test particle with energy, $E^2 =
0.895635~(\mu c^2)^2$, and angular momentum $L=3.54~G\mu
M/c$, where $\mu$ denotes the rest mass of the test
particle.  Although the $[+,-,+]$-region surrounds the
singularity (black dot), no LU region appears.}\\[1em] 
\noindent    
%%%%%%
\parbox[t]{2cm}{\bf FIG 2:\\~}\ \  
\parbox[t]{14cm} {(a) The $[+,-,+]$-regions (shaded) and LU
region (lightly shaded)  of the 2-Curzon spacetime  with two equal
masses, $M$,  located at $\pm 2 GM/c^2$ on the z axis
(black dots) and the bound region (dotted) of a test
particle with  the angular momentum $L=6.8~G\mu
M/c$ and energy, $E^2=0.88957~(\mu c^2)^2$
corresponding to $E_{\rm UPO}$.
 The LU region on the equatorial plane
intersects with the bound region (dark shaded).
(b)The Poincar\'e map of the bound orbits of
the test particle  in 2-Curzon spacetime
with the same parameters in (a).
We choose the equatorial plane as the Poincar\'e section. 
The initial
 momenta are all $p^\rho_0=
0$ except for the outermost one, which
is $p^{z}_{0}/p^{\rho}_{0}=0.01$. The initial positions are 
$z_0=0, \rho_0 = 3.13, 3.0 ({\rm chaotic}), 3.25 ({\rm chaotic}), 3.4 ,
2.83~GM/c^2$ from the innermost. 
Almost all of Tori are broken strongly. }\\[1em] 
%%%%%% 
\parbox[t]{2cm}{\bf FIG 3:\\~}\ \  
\parbox[t]{14cm} {The Poincar\'e map of the bound orbits of
the test particle  in 2-Curzon spacetime
with the same parameters as those in Fig.2, except for the
energy. We choose the equatorial plane
as the Poincar\'e section. 
We increase the energy  little by little up to the value of 
$E_{\rm UPO} (\approx 0.88957 \mu c^2)$
 as  $(E/\mu c^2)^2=$0.8892[(a)], 0.88935[(b)],
0.88957[(c)] in the Fig.2. 
The initial data for each tori in (a) are 
$p^\rho_0=0, 
z_0=0$ and $\rho_0 = 3.0, 3.1, 3.2, 3.3 ~GM/c^2$ from
the innermost.
The outermost torus corresponds to $p^{z}_{0}/p^{\rho}_{0}=0.01, 
z_0=0$ and $\rho_0 = 3.0~GM/c^2$.  In (b),  the two small
tori correspond to the initial conditions $p^\rho_0=
\pm 0.01 \mu c, 
z_0=0$ and $\rho_0 = 2.97~GM/c^2$.
Starting from the inside, the initial data for
the Poincar\'e maps are
$p^\rho_0=0, 
z_0=0$, and $\rho_0 = 3.0 ({\rm chaotic}), 2.97 ({\rm
chaotic}), 2.9, 3.15, 3.33~GM/c^2$. The outermost one
corresponds to $p^{z}_{0}/p^{\rho}_{0}=0.01,
z_0=0$ and $\rho_0 = 3.0~GM/c^2$. For (c), the initial
 momenta are all $p^\rho_0=
0$ except for the outermost one, which
is $p^{z}_{0}/p^{\rho}_{0}=0.01$. The initial positions are 
$z_0=0, \rho_0 = 3.13, 3.0 ({\rm chaotic}), 3.25 ({\rm chaotic}), 3.4 ,
2.83~GM/c^2$ from the innermost. 
 Tori are broken
more strongly as the energy increases. }\\[1em] 
\noindent   %%%%%%  
\parbox[t]{2cm}{\bf FIG 4:\\~}\ \ 
\parbox[t]{14cm} {The Poincar\'e map for the orbits of a
test particle in the 2-Curzon spacetime  with two equal
mass parameter $M$  located at $\pm 2 GM/c^2$ on z axis.
the energy and angular momentum are fixed as
$E^2=0.913~(\mu c^2)^2$ and $L=6.94~G\mu$, respectively
 so as to satisfy the condition
that only the orbit corresponding the most outer torus passes the
LU region.
 The equatorial plane is chosen as the
Poincar\'e section. The rests of the initial conditions are 
$p^\rho_0 =0, z_0=0, \rho_0=15.0, 10.0, 4.0~GM/c^2$ 
from inside torus. The
small torus around $(\rho , p^\rho)  \sim (3 ~GM/c^2, 0)$
corresponds to the orbit with the initial position of
$\rho_0=3.3 ~GM/c^2$. Only the torus which corresponds to
the orbit passing through the LU region is broken.}\\[1em]
 \noindent 
%%%%%%
\parbox[t]{2cm}{\bf FIG 5:\\~}\ \ 
\parbox[t]{14cm}
{Chaos in the 3-Curzon spacetime with three 
singularities of equal mass
parameter, $M$, located at the origin and at $\pm 10 GM/c^2$ on
the z axis. Two LU regions appear between
point singularities. The left figures [(a),(c),(e)] show
the LU region and the time evolution of the locus of the
bound orbit with  $E^2=0.76~(\mu c^2)^2$ and
$L=5.5~G\mu M/c$. The initial conditions for all figures 
are $p^z_0/p^\rho_0= 0.03, \rho_0=5.0~GM/c^2,
z_0=0$. The right figures
[(b),(d),(f)] are the Poincar\'e maps corresponding to the
orbits on the left side. The orbit initially departs from
the equatorial plane  and eventually approaches the LU
region. Then just after it crosses the LU region at the
proper time $\tau \approx 100337.5~GM/c^3~ (\sim 100
\times$ [the orbital period]), the torus begins to break
and the orbit becomes chaotic.}\\[1em]
 \noindent 
%%%%%%
\parbox[t]{2cm}{\bf FIG 6:\\~}\ \ 
\parbox[t]{14cm}
{Chaos in the 3-Curzon spacetime with three 
singularities of equal mass
parameter, $M$, located at the origin and at $\pm 10 GM/c^2$ on
the z axis (black dots).  The LU-region  is shown in (a) and (b)
with the locus of the bound orbits with the
same energy  $E^2=0.81~(\mu c^2)^2$ and angular momentum
$L=6.0~G\mu M/c$ but with different initial conditions
($\rho_0=10.0~GM/c^2, z_0=0$, and $p^\rho_0=0$ 
for (a) and $p^z_0/p^\rho_0= 0.6$ for (b) ). 
The Poincar\'e maps corresponding to those orbits 
are shown in (c). 
  The torus of the orbit which does not cross the LU region 
[(a)] is not broken, but  it is broken for 
 the orbit passing through the LU region [(b)].}\\[1em]
 \noindent 
%%%%%%
%%%%%%
\parbox[t]{2cm}{\bf FIG 7:\\~}\ \ 
\parbox[t]{14cm}
{Chaos in the 2-ZV spacetime with two singularities of equal mass
parameter $M$ ($m_1=m_2=M/\delta$) located one the lines between  $\pm 
GM/c^2$ and $\pm 3
GM/c^2$ on z axis (black solid lines). The parameter
$\delta$ is chosen to be 1.0, which means that each singularity
corresponds to the Schwarzschild black hole, then
$[+,-,+]$-regions do not appear around these
singularities. The LU and the bound
regions of the orbits are shown in (a), 
with  $E^2=0.90913~(\mu c^2)^2$,
$L=6.9~G\mu M/c$. In (a), the bound region certainly
intersects with the LU region.  
 The Poincar\'e map of the chaotic orbit in the bound region
 is shown in  (b). The rest of initial data for the orbit is 
$p^\rho_0=0, \rho_0=2.6~GM/c^2,z_0=0$.}\\[1em]
 \noindent 
 %%%%%% 
\parbox[t]{2cm}{\bf FIG 8:\\~}\ \  
\parbox[t]{14cm} {The  Lyapunov exponents of the orbits
with the same parameters as those in Figs.3(a) $\sim$
3(c) and with the initial conditions $p^\rho_0=0, 
\rho_0=3.0~GM/c^2,$ and $
 z_0=0$.  As we can see, 
the Lyapunov exponents for (b) and (c) converge to positive
values such that the larger value corresponds to the
larger energy of the particle.}\\[1em]  \noindent  
 %%%%%% 
\parbox[t]{2cm}{\bf FIG 9:\\~}\ \ 
\parbox[t]{14cm}
{The distribution of $<\kappa>$ on equatorial plane
 for 2-ZV spacetime with two singularities at $\pm 2 GM/c^2$ on
the z axis. We compared the $<\kappa>$  for the value of 
 $\delta$=0.65, 1.0, 10.0. The region where $<\kappa>~>0$
coincides with an LU region. For each $\delta$, $<\kappa>$
has a peak in an LU region. As the value of $\delta$ increases,
the peak becomes shorter and shorter.}\\[1em]
 \noindent 
%%%%%%
\parbox[t]{2cm}{\bf FIG 10:\\~}\ \ 
\parbox[t]{14cm}
{Chaos in the 2-Curzon spacetime without a reflection
symmetry. Two singularities with  mass parameters   $M$
and $0.5M$ are located  at $4GM/c^2$ and 
$-4GM/c^2$ on the z axis (black dots), respectively.
 The LU and the bound regions of the orbit with 
$E^2=0.865~(\mu c^2)^2$, $L=4.2~G\mu M/c$ are shown in (a),
while (b) shows its
Poincar\'e map. We chose the plane $z=2GM/c^2$ as a
Poincar\'e section. The initial conditions for the three
large tori are  $z_0=2.0~GM/c^2$ and
$(\rho_0, p^\rho_0) = (8.0~GM/c^2, 0.075 \mu c),
(8.0~GM/c^2, 0.05 \mu c), (7.5~GM/c^2, 0)$, and the three
small tori correspond to the orbit with the initial
data of $\rho_0=8.5~GM/c^2, z_0=2.0~GM/c^2$ and $p^\rho_0=-0.06
\mu c$.  The initial conditions for the chaotic
orbits are  $\rho_0=6.0~GM/c^2, z_0=2.0~GM/c^2$ and $
p^\rho_0=0$.
 Although the bound region does not intersect with the LU
region, strong chaos appears, as seen in (b).}\\[1em]
%%%%%%
\parbox[t]{2cm}{\bf Table 1:\\~}\ \ 
\parbox[t]{14cm}
{Comparison of $v_{\ast,max} \sqrt{<\kappa>_{max}}$ with  $\lambda$. The estimations are made for 2-ZV spacetime with two singularities at $\pm 2 GM/c^2$ on the z axis.
 $Q$ is a quadrupole moment and is given by $\delta$ as $\delta=(a) 0.65, (b) 1.0, (c)10.0$. In each case, $E$ and $L$ are determined so as to satisfy the condition, $E \sim E_{UPO}(L)$ and the bound region $D_{\rm eff}$ overlaps an LU region, as $(E/(\mu c^2)^2,~L/(G\mu M/c))=(a)(0.92713,\ 7.0), (b)(0.90913,\ 6.9), (c)(0.88959,\ 6.8)$. The rest of the initial condition are determined so that each geodesic becomes strongly chaotic as (a)$p^\rho_0=0, \rho_0=2.1~GM/c^2,z_0=0$, (b)$p^\rho_0=0, \rho_0=2.6~GM/c^2,z_0=0$, (c)$p^\rho_0=0, \rho_0=2.85~GM/c^2,z_0=0$.  For each $\delta$, $v_{\ast,max} \sqrt{<\kappa>_{max}}$ is in good agreement with the Lyapunov exponent $\lambda$. As the value of $\delta$ increases, both of the values become consistently smaller and smaller.}\\[1em] \noindent 
%%%%%%%
\end{document}